# Achievement of FCC specification in critical current density for Nb$_3$Sn superconductors with artificial pinning centers

X Xu[1], X Peng[2], J Rochester[3], M Sumption[3] and M Tomsic[2]

[1] Fermi National Accelerator Laboratory, Batavia, IL 60510, U.S.A

[2] Hyper Tech Research Incorporated, 539 Industrial Mile Road, Columbus, OH 43228, U.S.A

[3] Department of MSE, the Ohio State University, Columbus, OH 43210, U.S.A

E-mail: xxu@fnal.gov

**Abstract**

In this letter we demonstrate achievement of record non-Cu critical current density ($J_{c,non-Cu}$) in ternary, multifilamentary Nb$_3$Sn conductors by the introduction of artificial pinning centers (APC). In the past two years, we have made great progress in the development of APC Nb$_3$Sn wires. Recent resistivity vs magnetic field measurements confirmed the high upper critical field ($B_{c2}$) of ternary APC wires, which at 4.2 K was ~28 T, about 1-2 T higher than present state-of-the-art conductors. In addition to high $B_{c2}$, it was found that APC wires have noticeably higher Sn content in the Nb$_3$Sn layers as compared to standard wires. The $J_{c,non-Cu}$ values of the most-recent APC wires have met the $J_{c,non-Cu}$-$B$ specification required by the Future Circular Collider (FCC), with the best heat treatment leading to a $J_{c,nonCu}$ 29% higher than the FCC specification at 21 T. Microscopy analysis shows that the APC wires still have overly high residual Nb fractions due to too low of a Sn/Nb ratio, indicating that there is still great potential for further $J_{c,non-Cu}$ improvement. The development of APC wires is ongoing; this letter details some of the steps forward in the optimization and lays out a roadmap to push the APC wires towards practical, magnet-grade conductors.

**Keywords:** Nb$_3$Sn Superconductor, Artificial pinning center, $B_{c2}$, $J_c$.

**1. Introduction**

The proposed High-Energy Large Hadron Collider (HE-LHC) upgrade as well as Future Circular Collider (FCC) aims to push the energy frontier for high energy physics, and both require thousands of 16 T dipoles based on Nb$_3$Sn superconductors [1]. Design studies of such magnets have been pursued in earnest for a few years using Nb$_3$Sn conductor specifications that include a non-Cu critical current density versus field ($J_{c,non-Cu}$-$B$) curve which at 4.2 K and 16 T is 1500 A/mm$^2$ at a minimum [2,3]. Conductor $J_{c,non-Cu}$ is a critical factor for such large machines because building 16 T magnets with conductors of lower $J_{c,non-Cu}$ would require a larger coil size and thus significantly increase the overall project cost. The rod-restack-process (RRP®) wires, which are the present state-of-the-art Nb$_3$Sn conductors, meet the $J_{c,non-Cu}$ specification required by the High-Luminosity LHC (HL-LHC) upgrade with ~3% margin [4]. However, the FCC $J_{c,non-Cu}$ specification is 50% higher than the HL-LHC specification [1-3], which means that a ~45% improvement is still needed for state-of-the-art Nb$_3$Sn wires to fulfill the FCC requirement with a reasonable conductor acceptance fraction. This is a huge challenge considering that the record $J_{c,non-Cu}$ of Nb$_3$Sn wires has been at a plateau since the early 2000s [5] despite extensive efforts to break this barrier [6].

Introduction of artificial pinning centers (APC) to enhance flux pinning and thus $J_c$ has long been a "holy grail" in the Nb$_3$Sn area. Significant efforts had been made using various techniques since the 1980s but none succeeded in producing superior $J_c$ in wires [7-12] except proton or neutron irradiation [13,14] which is not practical for magnets. This was finally realized in 2014 [15] via the internal oxidation method, a technology that was used in Nb$_3$Sn tapes in the 1960s [16] but had been believed ineffective in Nb$_3$Sn wires after some unsuccessful attempts in the 2000s (e.g., [10,11]). About a decade later we revisited this idea and were able to find out the essentials for it to work in Nb$_3$Sn wires [15]. We first demonstrated using binary monofilaments the effects of this technique (e.g., significantly refined grain size, doubled layer $J_c$ at 12 T despite low irreversibility field – $B_{irr}$, shift of pinning force curve peak from $0.2B_{irr}$ to $0.34B_{irr}$, etc.) [15,17], and subsequently proposed applying it in the powder-in-tube (PIT) design in order to make multifilamentary wires [17,18]. Later binary multi-filamentary wires were fabricated by SupraMagnetics Inc. [19] and Hyper Tech Research Inc. [20], which confirmed the monofilament results. On the other hand, the low $B_{irr}$ seen in these binary APC wires [17,19,20] showed the need for ternary dopants. From 2017 several groups started to study the effect of ternary doping to APC wires, including Florida State University (FSU) [21] and University of Geneva [22], as well as a collaboration between Hyper Tech, Fermilab, and the Ohio State University (OSU) that began in earnest to develop ternary multifilamentary APC-PIT wires. As a surprising side result, FSU found in monofilaments grain refinement via Hf doping without internal oxidation, which also looks to be a promising method to improve Nb$_3$Sn $J_c$ [21].

The transport resistance vs field (*R-B*) and voltage vs current (*V-I*) tests up to 31 T in late 2018 [23] on the first set of ternary multifilamentary APC wires developed jointly by Hyper Tech, Fermilab, and OSU clearly demonstrated that the low $B_{irr}$ issue seen in binary APC wires had been solved by addition of Ta dopant, which was also seen in the monofilaments in [21] and [22]. In fact, the $B_{irr}$ and $B_{c2}$ values (4.2 K) of the APC wire with 1wt.% Zr (T3882-0.84mm-675°C/300h, denoted "APC-B" in [23]) were 26.8 and 27.6 T, respectively, ~2 T higher than the reference RRP® wire that is used for the HL-LHC. The measured $J_{c,non\text{-}Cu}$ (4.2 K, 16 T) of T3882-0.84mm-675°C/300h had reached a similar level to that of the reference RRP® wire. Scanning electron microscopy (SEM) analysis showed that the average fine-grain (FG) Nb$_3$Sn fraction in the filaments of T3882-0.84mm-675°C/300h was only 22%, while the residual Nb fraction was as high as 45%. Indeed, the FG $Nb_3Sn$ layer $J_c$ of T3882-0.84mm-675°C/300h was 2.5 times higher than that of the reference RRP® wire at 4.2 K, 16 T (4710 vs 1870 A/mm$^2$) [23]. This made it clear that there was still huge potential for further improvement of $J_{c,non-Cu}$ via precursor ratio optimization, which could allow more FG formation.

A second set of ternary APC wires had been developed, guided by the analysis on the first set of wires, and were tested in early 2019. The results, reported here, not only confirm the high $B_{irr}$ and $B_{c2}$ of ternary APC wires, but also demonstrate $J_{c,non-Cu}$ which has achieved (or surpassed) the FCC specification at high fields.

## 2. Experimental

### 2.1. Samples

The APC wire used here, T3912, is based on a 48/61-filament design (i.e., 48 $Nb_3Sn$ and 13 Cu filaments) with a Cu/non-Cu ratio of 1.3, fabricated at Hyper Tech using a Nb-1wt.%Zr-7.5wt.%Ta tube and a mixture of Cu, Sn, and $SnO_2$ powders. Compared with T3882 [23], the Sn/Nb and $SnO_2$/Nb ratios were increased in T3912. Most of the filaments of T3912 had sufficient oxygen. The wire was fabricated using a billet with a starting diameter of 19 mm, which was drawn to final wire diameters of 0.5, 0.71, and 0.84 mm, resulting in a total length of ~200 meters. Two commercial, state-of-the-art, $Nb_3Sn$ wires were tested together with the T3912 samples as references: one was a Ti-doped RRP® wire (billet number 00076) with 0.85mm diameter and 108/127 subelements, and the other was a standard PIT wire (billet number 31284) with 0.78mm diameter and 192/217 filaments; both wires were manufactured by Bruker EST and used for HL-LHC.

All wires were heat treated under vacuum. The APC wires were reacted at 675°C, 685°C, and 700°C for various durations (specified below), all with a ramp rate of 30°C/h without intermediate steps. The RRP® wire used the HL-LHC HT protocol: 210°C/48h + 400°C/48h + 665°C/75h [4]. The standard PIT wire used a two-stage HT 630°C/100h + 640°C/50h in order to maximize its $J_{c,non-Cu}$ while maintaining high residual resistivity ratio (RRR) [24].

*2.2. Measurements*

Two types of transport tests were performed on these wires at the National High Magnetic Field Laboratory (NHMFL): (1) *R-B* measurements using a sensing current of 31.6 mA, allowing determinations of $B_{c2}$ and $B_{irr}$, (2) *V-I* measurements at various fields, giving critical current ($I_c$) values. All tests were at 4.2 K in a 31T magnet with fields perpendicular to wire axes. Great care was taken to ensure that each sample was centered within the magnet. The *R-B* tests were performed on samples 15mm in length with a voltage tap spacing of 5 mm. Since five samples could be mounted together in the holder for each run, we always included one reference wire in order to give a double check on the results. The *V-I* measurements used the $I_c$ transport test system from the Applied Superconductivity Center (ASC) at the NHMFL (including the $I_c$ rig, sample holders, data acquisition system, testing program, etc.), and were performed on straight samples 35mm in length with a voltage tap separation of 6 mm. A criterion of 0.1 µV/cm was used to determine the $I_c$ values. The compositions of the $Nb_3Sn$ layers were measured using energy-dispersive spectroscopy (EDS): 25kV voltage was used, with a collection rate above 40000 counts per second, and for each spot data were collected for 30 seconds.

**3. Measurement results**

The *R-B* curve of T3912 (0.84mm diameter given a HT of 675°C/384h) is shown in Figure 1, along with those of T3882-0.84mm-675°C/300h [23] as well as the reference wires. Below we take the fields at 10% and 90% of the normal state resistances as $B_{irr}$ and $B_{c2}$, respectively.

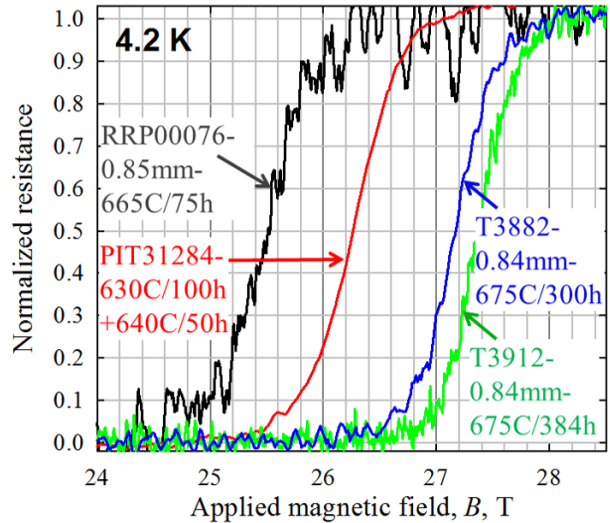

Figure 1. The 4.2K *R-B* curves of the reference wires as well as T3882-0.84mm-675°C/300h [23] and T3912-0.84mm-675°C/384h.

The $B_{irr}$ and $B_{c2}$ of the RRP® wire were ~25 and ~26 T, respectively, which are typical for RRP® wires reacted at 665°C. The standard PIT wire had $B_{irr}$ and $B_{c2}$ of 25.8 and 26.7 T, respectively, which are similar to what Godeke measured for PIT wires reacted at 675°C [25] and are nearly 1T higher than those of the RRP® wire. The 4.2K $B_{irr}$ and $B_{c2}$ of T3912-0.84mm-675°C/384h were 27 and 27.8 T, respectively, over 1T higher than those of the standard PIT wire. These results confirm our previous measurements in late 2018 [23] showing that the ternary APC wires have higher $B_{irr}$ and $B_{c2}$ than the standard PIT and RRP® conductors.

The measured $J_{c,non-Cu}$ values of the reference wires and T3882-0.84mm-675°C/300h [23] as well as T3912 given various heat treatments are shown in Figure 2, along with the FCC $J_{c,non-Cu}$-
6

$B$ specification curve that is generated using the equations and parameters given in [3]. Here it is worth mentioning that design of a magnet with 16T bore field (which leads to a slightly higher peak field on the conductors [3]) in fact requires the use of $J_{c,non-Cu}$ at a higher field (e.g., ~19 T for a 14% margin [3] along the peak field load line) – indeed, a calculation of conductor temperature margin shows that the 19T $J_{c,non-Cu}$ is more relevant than the 16T $J_{c,non-Cu}$; therefore, the 19T $J_{c,non-Cu}$ values will also be reported below. The 16T $J_{c,non-Cu}$ values of the RRP® and PIT wires were 1090 and 1005 A/mm², and the 19T values were 395 and 410 A/mm², respectively. These are not the highest values for top-performing RRP® and PIT wires (which are ~1300 A/mm² at 16 T [2]), but are in their typical $J_{c,non-Cu}$ spectrums. Despite lower $J_{c,non-Cu}$ at low fields, PIT wires outperform RRP® at high fields (19 T and above) due to higher $B_{irr}$, which has also been seen previously [2]. Because the $I_c$ values of the APC wires with 0.84mm diameter could not be obtained at low fields due to quenching during the $V$-$I$ tests (which could be due to large subelement size $D_s$ and a bit too long reaction times), the $J_{c,non-Cu}$-$B$ data are fitted with a commonly-used two-component pinning force model, which has been shown by a number of studies to work well for both standard and APC wires [9,19,23]. As can be seen in Figure 2, the fittings to the data are good. The $J_{c,non-Cu}$ of T3912 at 0.71mm diameter (with $D_s$ of 70 μm) after a HT of 700°C/71h was 3% lower than the FCC specification at 16 T, but surpassed the latter at 19 T and above. The RRR of this wire was 85 – improvement of RRR requires further improvement of wire quality as well as HT optimization studies to prevent too long reaction time. With the same HT temperature of 700°C, an increase of the wire diameter to 0.84 mm (with $D_s$ of 83 μm) led to only slightly higher $J_{c,non-Cu}$ (by <5%). On the other hand, lower HT temperatures led to substantially higher $J_{c,non-Cu}$: e.g., reducing HT to 685°C increased the $J_{c,non-Cu}$ to 730 A/mm² at 19 T (i.e., 13% higher than the FCC specification) and extrapolated 1600 A/mm² at 16 T. Further reducing HT to 675°C led to even higher $J_{c,non-Cu}$: e.g., 380 A/mm$^2$ at 21 T (i.e., 29% higher than the FCC specification), and extrapolated values of 780 A/mm$^2$ at 19 T and 1740 A/mm$^2$ at 16 T, the latter being 16% higher than the FCC specification and ~30% higher than the RRP® record.

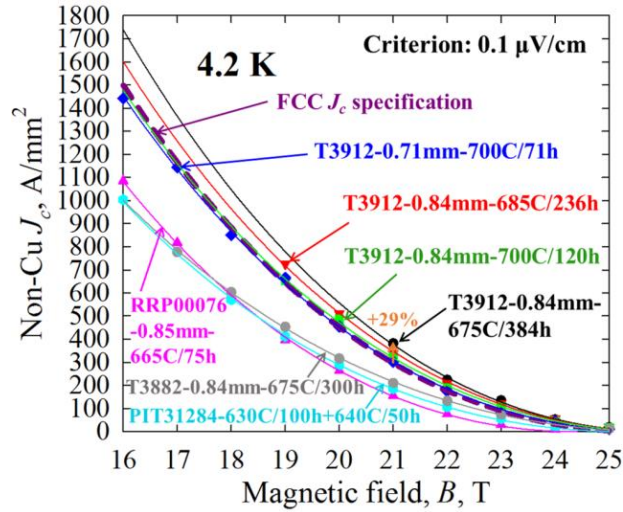

Figure 2. Non-Cu $J_c$s of the reference and the APC wires, as well as the FCC $J_{c,non-Cu}$-$B$ specification.

## 4. Discussions

In addition to the high $B_{irr}$ and $B_{c2}$, we found that APC wires also had higher Sn content in the FG layers as compared to that in conventional wires. The Sn concentrations in Nb$_3$Sn of T3912 as well as those of several other wires (all with 48/61 design) with different combinations of Zr, Ta, and O, as shown in Table 1, were measured and are shown in Figure 3. For all of the samples, the first EDS spots were in the coarse-grain (CG) Nb$_3$Sn regions (which are known to have Sn content around 25 at.% [26]), and the last spots were close to the FG/Nb interfaces, while all the other spots were in the FG layers. In Nb$_3$Sn wires, only the FG layers can carry supercurrents, while the CG cannot. It can be seen that T3736, T3814, and T3843 generally had Sn concentrations in FG between 22.5 and 24 at.%, which are typical for standard Nb$_3$Sn wires [26]. On the other hand, T3739 and T3912 had Sn concentrations in FG between 24 and 25 at.%, close to stoichiometry. By comparing these wires, it can be clearly seen that only the internally oxidized samples (Zr+O) had the highest Sn content in FG. To verify the Sn content results, several APC and standard PIT and tube type wires given the same HT (675°C/120h) were sent as blind samples to Ian Pong at Lawrence Berkley National Laboratory (LBNL) for EDS verification measurements. The results of detailed line-scan measurements by Pong confirmed that the Sn concentrations of the APC wires were 1-1.5 at.% higher than those of standard wires [27]. The enhanced Sn content in the APC wires is possibly due to the reduced reaction rate at the Nb$_3$Sn/Nb interface [17], which, according to a diffusion reaction theory for Nb$_3$Sn composition [6,28], benefits the Sn content of a Nb$_3$Sn layer.

Table 1. Summary of the wires for EDS studies

| Wire | Diameter, mm | Tube | Core | Chemistry |
| --- | --- | --- | --- | --- |
| T3736 | 1.0 | Nb-7.5wt.%Ta | Cu-clad Sn rod | With Ta, but no Zr or O |
| T3739 | 0.84 | Nb-1wt.%Zr | Mixture of Cu, Sn, and SnO$_2$ powders | With Zr and O, but no Ta |
| T3814 | 0.84 | Nb-1wt.%Zr-7.5wt.%Ta | Mixture of Cu and Sn powders | With Zr and Ta, but no O |
| T3843 | 0.84 | Nb-7.5wt.%Ta | Mixture of Cu, Sn, and SnO$_2$ powders | With Ta and O, but no Zr |

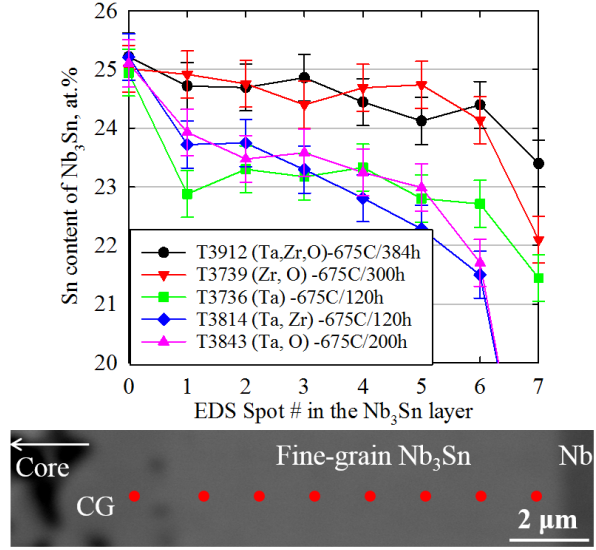

Figure 3. Sn concentrations in $Nb_3Sn$ layers of the selected wires and a schematic showing the EDS spots.

The $B_{irr}$ and $B_{c2}$ of $Nb_3Sn$ depend on both Sn content and doping. Ti and Ta dopants may have different influences on $B_{c2}$ and also on Sn content due to possible different site occupancy, which was studied in [29] and recently [30]. Here we compare standard PIT and our APC wires, all based on Ta doping. It is well known (e.g., [6,26,29]) that for ternary $Nb_3Sn$, $B_{c2}$ increases with Sn content. Therefore, the higher Sn content in FG $Nb_3Sn$ of the ternary APC wires could be why they have higher $B_{irr}$ and $B_{c2}$ than standard PIT wires.

SEM images of cross sections of T3912-0.71mm-700°C/71h and T3912-0.84mm-675°C/384h are shown in Figure 4. From these images the average area fractions of FG, residual Nb, and CG in filaments of both wires were calculated and are shown in Table 2, along with those of T3882-0.84mm-675°C/300h [23] as well as standard PIT conductors. From Table 2, the FG fractions of both T3912 wires are nearly 50% higher than that of the T3882-0.84mm-675°C/300h: this is the primary reason for the much higher $J_{c,non-Cu}$ of T3912 in comparison to T3882; a secondary reason is an improvement of wire uniformity (T3882 had a number of filaments with insufficient oxygen [23]). Since T3912-0.84mm-675°C/384h has lower FG fractions than T3912-0.71mm-700°C/71h, its higher $J_{c,non-Cu}$ is most likely due to higher flux pinning as a result of the lower HT temperature. The residual Nb fractions of both T3912 wires (34%) are still much higher than those of standard PIT wires (typically 25% or below [24]). A closer analysis of the recipe of T3912 reveals that this is because it has too low of a Sn/Nb ratio. This means that the APC wires still have the potential to form more FG by converting more Nb by properly tweaking the recipe design, which is expected to lead to higher $J_{c,non-Cu}$.

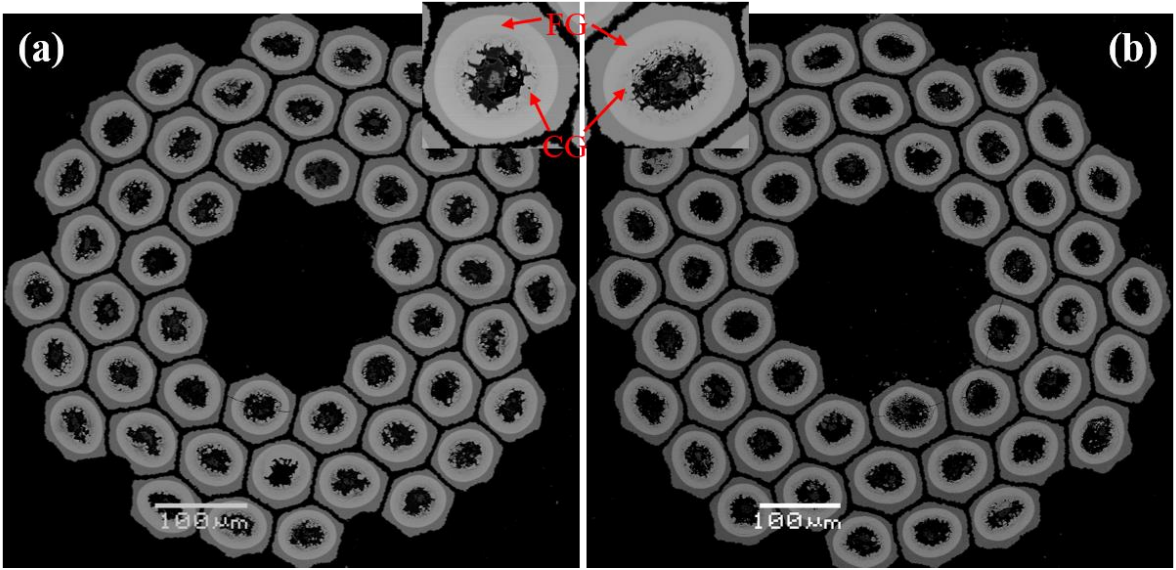

Figure 4. SEM images of cross sections for (a) T3912-0.71mm-700°C/71h, and (b) T3912-0.84mm-675°C/384h.

Table 2. Area fractions in filaments of T3882 and T3912 wires as well as standard PIT wires

| Areas | T3882-0.84mm-675°C/300h [23] | T3912-0.71mm-700°C/71h | T3912-0.84mm-675°C/384h | Standard PIT [24] |
|---|---|---|---|---|
| Fine-grain A15 | 22% | 33% | 31% | 40% |
| Residual Nb | 45% | 34% | 34% | 25% |
| Coarse-grain A15 | 13% | 14% | 16% | 13% |

Despite achievement of the FCC $J_{c,non-Cu}$ specification, there is still much work to be done. First, it is necessary to continue to optimize the wire recipe, which on one hand may lead to better Nb/Sn ratio for a fuller reaction and thus higher $J_{c,non-Cu}$, and on the other hand will also likely further improve wire quality and thus RRR, considering that recipe optimization has been the primary contributor to the improvement of APC wire quality (e.g., filament shape) in the past two years. Second, it is important to continue to optimize the HT of APC wires because a lower HT temperature benefits $J_{c,non-Cu}$ and proper reaction times are required to retain high RRR. Third, exploration of some wire design variants that have been demonstrated in standard PIT wires could be very helpful in improving performance of APC wires, such as the use of round filaments instead of the present hexagonal ones (this improves $J_{c,non-Cu}$ because it allows more Nb to be converted into Nb$_3$Sn without poisoning the RRR) and the use of a bundle barrier (which is effective in preserving high RRR) [31]. Next, it is necessary to continue the fabrication and optimization of 217-filament wires, which are important for reduction of $D_s$ and thus improvement of conductor stability. We have fabricated a wire with 217 filaments using a Nb-0.6wt.%Zr-7.5wt.%Ta tube, which was drawn to 0.71mm diameter (with $D_s$ of 35 μm) without wire breakage. A next logical step would be fabrication of wires with 217 filaments using Nb-1wt.%Zr-7.5wt.%Ta tubes for higher $J_{c,non-Cu}$. Here it is worth mentioning that APC wires demonstrate good drawability: all the APC wires of the past 23 billets (including 13 using Nb-0.6wt.%Zr-7.5wt.%Ta tubes and 10 using Nb-1wt.%Zr-7.5wt.%Ta tubes) were drawn to 0.5-0.84mm diameters in single pieces without any wire breakage. In parallel with the above research and development tasks, there are other tasks which are needed in order to push APC wires towards practical, magnet-grade conductors, such as process improvement and increased billet size, etc., as well as evaluation of electro-mechanical properties (e.g., uniaxial tensile and transverse pressure tests), which are important for magnet applications.

## 5. Conclusions

Our results reported here on multifilamentary APC $Nb_3Sn$ conductors confirm the high $B_{irr}$ and $B_{c2}$ of ternary, internal oxidation route APC wires (27-28 T at 4.2 K), which are 1-2 T higher than RRP® and standard PIT wires. This is because APC wires have significantly higher Sn content in the $Nb_3Sn$ layer as compared to those without internal oxidation. The $J_{c,non-Cu}$ values of these recent APC wires have achieved the FCC $J_{c,non-Cu}$-$B$ specification, and a HT at 675°C led to a $J_{c,non-Cu}$ 29% higher than the FCC specification at 21 T. Analysis of SEM images showed that the FG $Nb_3Sn$ fractions have increased dramatically, but are still below what we estimate is possible for properly optimized chemistries. Finally, some work remains to be done in order to further improve performance of APC wires and push them towards practical conductors, and we see a potentially viable path to make magnet-grade, long-length APC $Nb_3Sn$ conductors.

**Acknowledgements**

This work is supported by the Laboratory Directed Research and Development (LDRD) program of Fermilab and Hyper Tech SBIR DE-SC0013849 and DE-SC0017755 by US Department of Energy. Some tests were performed at the NHMFL, which is supported by National Science Foundation Cooperative Agreement No. DMR-1644779 and the State of Florida. The measurements at the NHMFL were greatly helped by Jan Jaroszynski, Griffin Bradford, and Yavuz Oz. We are grateful to ASC of the NHMFL for letting us use the $I_c$ transport test system. We thank Ian Pong from LBNL for supplying the RRP® and PIT wires that are used as references in this work, and for the independent EDS studies that confirmed our results. This work is performed under the auspices of U.S. Magnet Development Program.